\documentclass[pra,twocolumn,groupedaddress,showpacs,floatfix]{revtex4}

\usepackage[usenames, dvipsnames]{color}
\usepackage{graphicx}
\usepackage{bm}
\usepackage{amsmath}
\usepackage{amsfonts}
\usepackage{amssymb}
\usepackage{latexsym}
\usepackage{cancel}
\usepackage{bbm}
\usepackage{hyperref}

\renewcommand{\>}[0]{\rangle}
\newcommand{\<}[0]{\langle}

\newcommand{\proj}{\mathbb{P}}

\begin{document}

\title{Atomic cluster state build up with macroscopic heralding}
\author{Jeremy Metz,$^{1}$ Christian Sch\"on,$^{1}$ and Almut Beige$\,^{2}$}
\affiliation{$^1$Blackett Laboratory, Imperial College London, Prince Consort Road, London SW7 2BZ, United Kingdom \\
$^2$The School of Physics and Astronomy, University of Leeds, Leeds LS2 9JT, United Kingdom}

\date{\today}

\begin{abstract}
We describe a measurement-based state preparation scheme for the efficient build up of cluster states in atom-cavity systems. As in a recent proposal for the generation of maximally entangled atom pairs [Metz {\em et al.}, Phys. Rev. Lett. {\bf 97}, 040503 (2006)], we use an {\em electron shelving} technique to avoid the necessity for the detection of single photons. Instead, the successful fusion of smaller into larger clusters is heralded by an easy-to-detect macroscopic fluorescence signal. High fidelities are achieved even in the vicinity of the bad cavity limit and are essentially independent of the concrete size of the system parameters.
\end{abstract}

\maketitle

\section{Introduction} \label{sec:intro}

In 2001 Raussendorf and Briegel pointed out that certain highly entangled states present an innovative approach to quantum computing \cite{oneway}. The attractiveness of these so-called cluster states \cite{cluster} arises from the fact that they can be grown off-line in a probabilistic fashion. Afterwards, a so-called one-way quantum computation can be carried out without having to create additional entanglement. Any quantum algorithm can then be performed using only single-qubit rotations and single-qubit measurements. Scalable fault-tolerant one-way computation is possible, provided the noise in the implementation is below a certain threshold \cite{nielsen,nielsen2}. For example, Raussendorf {\em et al.} \cite{rob2} recently introduced a fault-tolerant three dimensional cluster state quantum computer based on methods of topological quantum error correction. Other authors identified highly efficient cluster state purification protocols \cite{pachos,rob}.

A very efficient way to create a cluster state of a very large number of atoms with very few steps is to employ cold controlled collisions within optical lattices with one atom on each site \cite{Jaksch}. Using this approach, Mandel {\em et al.} \cite{Mandel} already created cluster state entanglement and reported the observation of coherence of an atom de-localised over many sites. Unfortunately, single-qubit rotations cannot be easily realised, since laser fields applied to one atom generally affect also its neighbours. To facilitate one-way quantum computing in optical lattices several schemes have been proposed for the realisation of single-qubit rotations without having to address the atoms individually \cite{Scheune,Zoller,Joo}.

However, higher fidelities can be obtained using a measurement-based cluster state growth approach. An example is the linear optics proposal by Browne and Rudolph \cite{Rudolph}. Using linear optics, a four-photon cluster state has already been generated in the laboratory \cite{Walther}. Currently, the scalability of this approach is hampered by the lack of reliable photon storage. To overcome this and the above mentioned addressability problem in optical lattices, quantum computing architectures have been proposed using hybrid systems based on atomic {\em and} photonic qubits \cite{Cabrillo,Plenio,Kok,Lim,Lim2}. To create entanglement between distant atoms, the atoms are operated as sources for the generation of single photons on demand followed by photon pair measurements in a carefully chosen basis. The main limitation of these two approaches lies in the difficulty of detecting single photons.

\begin{figure}
\begin{minipage}{\columnwidth}
\begin{center}
\resizebox{\columnwidth}{!}{\rotatebox{0}{\includegraphics {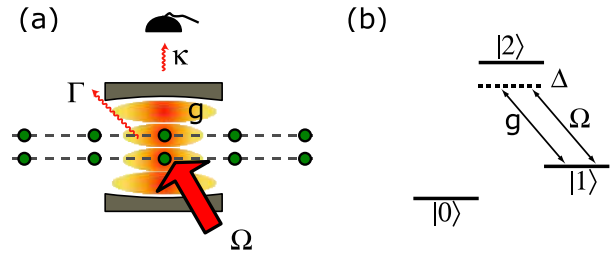}}}
\end{center}
\vspace*{-0.5cm} 
\caption{(Colour online) (a) Experimental setup for the cluster state growth of atomic qubits. Fusing two clusters requires placing an atom from each cluster into the resonator, where they are illuminated by a laser field. (b) Level scheme of a single atom in the cavity. Both atoms should experience similar coupling constants.} \label{fig:setup}
\end{minipage}
\end{figure}

Here we avoid the necessity of detecting single photons. Instead the successful fusion of two smaller clusters into a larger one is heralded by a {\em macroscopic} fluorescence signal. The experimental setup is shown in Fig.~\ref{fig:setup}(a). Suppose two atoms are placed into the antinodes of an optical cavity and a laser with Rabi frequency $\Omega$ and detuning $\Delta$ is applied. It drives the 1--2 transition of each atom as shown in Fig.~\ref{fig:setup}(b). The same  transition should also interact with the resonator field with the atom-cavity coupling constant $g$ and detuning $\Delta$. We denote the decay rate of a single photon through the cavity mirrors by $\kappa$ \cite{kappa} and $\Gamma$ is the spontaneous decay rate of the excited atomic state. Here we are especially interested in the parameter regime where
\begin{eqnarray} \label{eqn:conditions}
\Delta &\gg & \Omega,\, g,\, \kappa, \, \Gamma \, .
\end{eqnarray}
A detector constantly monitors the emission of photons through the cavity mirrors.  

As in Refs.~\cite{MTB-PRL,mebe}, we assume that both atoms experience similar interactions. Here the state $|0\>$ is decoupled from any dynamics of the system. As we see below, there are therefore three distinct fluorescence levels in the leakage of photons through the cavity mirrors. These are similar to the discrete fluorescence levels of two dipole interacting three-level atoms, which exhibit macroscopic quantum jumps \cite{jumps,jumps2,jumps3}. The lack of cavity photons indicates that both atoms are in $|0 \rangle$. The emission of cavity photons at a lower intensity level indicates that one atom is in $|0 \rangle$ and one atom is in $|1 \>$ without revealing which one, while cavity photons at a maximum rate indicate that the atoms are in $|11 \>$. In fact, the observation of the cavity fluorescence implements a probabilistic parity measurement with the projections
\begin{eqnarray} \label{eqn:parity}
\proj_{00} & \equiv & |00\>\<00| \, , \nonumber \\
\proj_{01}+ \proj_{10} &\equiv & |01 \> \< 01| + |10 \> \< 10| \, , \nonumber\\
\proj_{11} &\equiv & |11\>\<11| \, .
\end{eqnarray}
The successful projection of the atoms onto the subspace spanned by the states $|01 \>$ and $|10 \>$ can be used to create entanglement \cite{Franson,Molmer2,Molmer3}. For example, 
\begin{eqnarray} \label{eqn:parity_example1}
&& \hspace*{-2cm} (\proj_{01}+ \proj_{10}) \big[ {\textstyle \frac{1}{2}}(|0 \> + |1 \>) \otimes (|0 \rangle + |1 \rangle) \big] \nonumber \\
&= & {\textstyle \frac{1}{2}} (|01 \> + |10 \>) \, . 
\end{eqnarray}
However, it can also be used to generate entanglement between atoms without destroying any previous entanglement of these atoms with other atoms. For example, the projection $\proj_{01}+ \proj_{10}$ applied to atoms 2 and 3 obtained from two different Bell pairs, 
\begin{eqnarray} \label{eqn:parity_example2}
~~~ && \hspace*{-2cm} (\proj_{01} + \proj_{10})^{(2,3)}  
\big[ {\textstyle \frac{1}{2}}(|01 \> + |10 \>) \otimes (|01 \> + |10 \>) \big] \nonumber \\
&= & {\textstyle \frac{1}{2}} (|0101 \> + |1010 \>) \, ,
\end{eqnarray}
results in the generation of a four-atom GHZ state. Browne and Rudolph moreover showed that the measurement (\ref{eqn:parity}) enables the fusion of two smaller cluster states into one larger one with a success rate of $50 \, \%$ \cite{Rudolph}. It can therefore be used for the sequential build up of large cluster states. Detailed analyses on the scalability of related probabilistic cluster state growth schemes can be found for example in Refs.~\cite{Kok,Lim2,Nielsenxxx,Earl,Gross}.

Achieving high fidelities is possible, even when using moderate atom-cavity systems with relatively large spontaneous decay rates. The reason is that the qubits are encoded in long-living atomic ground states. Moreover, as in Refs.~\cite{MTB-PRL,mebe,Molmer3,Molmer2}, we can allow for single atom-cooperativity parameters $C$,
\begin{eqnarray} \label{eqn:coop}
C &\equiv & \frac{g^2}{\kappa \Gamma} \, ,
\end{eqnarray}
of the order of one and larger, as they are currently becoming available in the laboratory \cite{Chapman,Kuhn,Kimble, Rempe,Meschede3,Chapman07}. For $C=1$ and when using a perfect single photon detector, we show that it is possible to achieve fidelities above $0.88$. Lower photon detector efficiencies $\eta$ require larger $C$'s. For example, if $\eta = 0.2$ the cooperativity parameter $C$ should be 5 or larger. 

The above described distinct fluorescence levels occur in the emission from the cavity mode for a very wide range of experimental parameters. The performance of the proposed state preparation scheme is therefore essentially independent of the concrete size of the system parameters. To illustrate this we show that high fidelity parity measurements are possible even when the two atoms experience coupling constants differ from each other by up to $30 \, \%$. Once a cluster state has been built, performing a one-way quantum computation requires only single-qubit rotations and measurements as they are routinely used in ion trap experiments \cite{Blattt2,wineland}. More specifically, read out measurements are performed via the creation of macroscopic fluorescence signals and too have a very high accuracy even when using finite efficiency photon detectors \cite{BeHe}. 

It is experimentally feasible to trap two atoms fairly accurately in different antinodes of the cavity field. The efficiency of cavity cooling has recently been demonstrated by Nu{\ss}mann {\em et al.} \cite{cooling}. Domokos and Ritsch showed that it is possible to take advantage of cavity-mediated forces to keep the atoms predominantly at positions with maximum atom-cavity couplings \cite{Domokos}. A disadvantage of the proposed state preparation scheme lies in the necessity to move atoms in an out of an optical cavity. The so-called shuttling of atoms \cite{Hensinger} is relatively time consuming and limits the efficiency with which one can build large cluster states. However, its feasibility has already been demonstrated by several groups who combined for example atom trapping \cite{Chapman07,Meschede,Meschede2,Kimble} or ion trapping \cite{Blattt,Meschede0} technology with optical cavities. New perspectives arise from the development of atom-cavity systems mounted on atom chips \cite{Meschede3,Meschede4}. 

There are six sections in this paper. In Section \ref{sec:theory} we describe the setup shown in Fig.~\ref{fig:setup} and derive its effective dynamics. In Section \ref{sec:distinct} we discuss the nature of the three distinct levels in the fluorescence through the cavity mirrors and describe how to exploit these for the implementation of the probabilistic parity measurement (\ref{eqn:parity}). In Section \ref{sec:parity} we analyse the performance of the proposed protocol. In Section \ref{sec:cluster} we review the cluster state build up with parity checks. Finally, we summarise our findings in Section \ref{sec:conc}. Some mathematical details are given in Appendix \ref{app:probs}.

\section{Theoretical model} \label{sec:theory}

In this section we use the quantum jump approach \cite{Hegerfeldt,Molmer,Carmichael} to obtain an effective theoretical model for the description of the atom-cavity system in Fig.~\ref{fig:setup}.

\subsection{The no-photon evolution}

Proceeding as in Ref.~\cite{Hegerfeldt}, i.e.~assuming rapidly repeated environment-induced measurements and starting from the total Hamiltonian for the atom-cavity system and the surrounding free radiation fields, one can show that the (unnormalised) state of the system under the condition of {\em no} photon emission within $(0,t)$ can be written as
\begin{eqnarray} \label{eqn:Ucond}
|\psi^0(t) \> &=& U_{\rm cond}(t,0) \, |\psi_0 \> \, . 
\end{eqnarray}
Here $|\psi_0 \>$ is the state of the system at $t=0$ and $U_{\rm cond}(t,0)$ is the no-photon evolution operator. The corresponding conditional Hamiltonian equals, with respect to an appropriately chosen interaction picture, 
\begin{eqnarray} \label{eqn:Hcond}
H_{\rm cond} &=& \sum_{i=1,2} {\textstyle \frac{1}{2}} \hbar \Omega \, \big[ \, |1 \>_{ii} \< 2| + |2\>_{ii} \< 1| \, \big] \nonumber \\
&& + \sum_{i=1,2} \hbar g \, \big[ \, |1\>_{ii} \< 2| \, b^\dagger + |2 \>_{ii} \< 1| \, b \, \big] \nonumber\\
&& + \sum_{i=1,2} \hbar \Big( \Delta - {\textstyle \frac{{\rm i} }{ 2}} \Gamma \Big) \, |2 \>_{ii} \< 2| - {\textstyle \frac{{\rm i} }{ 2}} \hbar \kappa \, b^\dagger b \, . ~~~
\end{eqnarray}
The non-Hermitian terms in the last line of this equation damp away population in states that can cause an emission. After renormalisation of the state vector $|\psi^0(t) \>$, this results in a relative increase in population of states with a lower spontaneous decay rate. In this way, the quantum jump approach takes into account that the observation of no photons reveals information about the system. It gradually reveals that the system is more likely to be in a state where it cannot emit. 

In the following, we decompose $|\psi^0 \>$ in Eq.~(\ref{eqn:Ucond}) as 
\begin{eqnarray} \label{eqn:PsiDecomp}
|\psi^0 \> &=& \sum_{j,k=0}^2 \, \sum_{n=0}^\infty c_{jk;n} \, |jk;n \> \, , 
\end{eqnarray}
where $c_{jk;n}$ is the amplitude of the state $|jk;n \>$ with the first atom in $|j \>$, the second atom in $|k \>$ and $n$ photons in the cavity mode. According to the Schr\"odinger equation, the evolution of these coefficients is given by
\begin{eqnarray} \label{eqn:coeffgeneral}
\dot{c}_{jk;n} &=& - \frac{\rm i}{\hbar} \, \< jk;n | H_{\rm cond} | \psi^0 \> \, .
\end{eqnarray}
Writing out these equations and using Eq.~(\ref{eqn:conditions}) we find that the coefficients of states with population in $|2 \>$ change on a much faster time scale than the coefficients of atomic ground states. We can therefore eliminate them adiabatically from the system's evolution. Doing so, and setting their derivative equal to zero, we obtain
\begin{eqnarray} \label{eqn:coeffexatom}
	c_{02;n} &=& \frac{\rm i}{4 \Delta^2} \big(2 {\rm i} \Delta - \Gamma - n \kappa \big) \nonumber \\
	         && \times \big[ \Omega \, c_{01;n} + 2 \sqrt{n+1} g \, c_{01;n+1} \big] \, , \nonumber\\
	c_{20;n} &=& \frac{\rm i}{4 \Delta^2} \big(2 {\rm i} \Delta - \Gamma - n \kappa \big) \nonumber \\
	         && \times \big[ \Omega \, c_{10;n} + 2 \sqrt{n+1} g \, c_{10;n+1} \big] \, , \nonumber\\
	c_{12;n} = c_{21;n} &=& \frac{\rm i}{4 \Delta^2} \big(2 {\rm i} \Delta - \Gamma - n \kappa \big) \nonumber \\
	         && \times \big[ \Omega \, c_{11;n} + 2 \sqrt{n+1} g \, c_{11;n+1} \big] \, , \nonumber\\
	c_{22;n} &=& \frac{1}{4 \Delta^2} \big[ \Omega^2 \, c_{11;n} + 4 \sqrt{n+1} \Omega g \, c_{11;n+1} ~~ \nonumber \\
	         && + 4\sqrt{(n+1)(n+2)} g^2 \, c_{11;n+2} \big] 
\end{eqnarray}
up to second order in $1/\Delta$, given that most of the population remains in the atomic ground states. Substituting Eq.~(\ref{eqn:coeffexatom}) into the differential equation (\ref{eqn:coeffgeneral}), we then find that 
\begin{widetext}
\begin{eqnarray} \label{eqn:coeffgroundatom}
	{\dot{c}}_{00;n} &=& - {\textstyle \frac{1}{2}} n \kappa \, c_{00;n} \, , \nonumber\\
	{\dot{c}}_{01;n} &=& \frac{\Omega}{8 \Delta^2} \big(2 {\rm i} \Delta - \Gamma - n \kappa \big) \big[ \Omega \, c_{01;n} +
		 2 \sqrt{n+1} g \, c_{01;n+1} \big] + \frac{\sqrt{n} g}{4 \Delta^2}  \big( 2 {\rm i} \Delta - \Gamma - (n-1)\kappa \big) \big[ \Omega \, c_{01;n-1} 
		 +2 \sqrt{n} g c_{01;n} \big] \nonumber \\
		 && - {\textstyle \frac{1}{2}} n \kappa \,  c_{01;n} \, ,
	\nonumber\\
	{\dot{c}}_{10;n} &=& \frac{\Omega}{8 \Delta^2} \big(2 {\rm i} \Delta - \Gamma - n \kappa \big) \big[ \Omega \, c_{10;n} +
		 2 \sqrt{n+1} g \, c_{10;n+1} \big] + \frac{\sqrt{n} g}{4 \Delta^2}  \big( 2 {\rm i} \Delta - \Gamma - (n-1)\kappa \big) \big[ \Omega \, c_{10;n-1} 
		 +2 \sqrt{n} g c_{10;n} \big] \nonumber \\
		 && - {\textstyle \frac{1}{2}} n \kappa \,  c_{10;n} \, ,
	\nonumber\\
	 {\dot{c}}_{11;n} &=& \frac{\Omega}{4 \Delta^2} \big(2 {\rm i} \Delta - \Gamma - n \kappa \big) \big[ \Omega \, c_{11;n} +
		 2 \sqrt{n+1} g \, c_{11;n+1} \big] + \frac{\sqrt{n} g}{2 \Delta^2} \big( 2 {\rm i} \Delta - \Gamma - (n-1)\kappa \big) \big[ \Omega \, c_{11;n-1} 
		 +2 \sqrt{n} g c_{11;n} \big] \nonumber \\
		 && - {\textstyle \frac{1}{2}} n \kappa \,  c_{11;n} 
\end{eqnarray} 
\end{widetext}
up to second order in $1/\Delta$. Given the parameter regime in Eq.~(\ref{eqn:conditions}), these differential equations contain two very different time scales. Since $\kappa$ is much larger than all other frequencies that scale as $1/\Delta$ or $1/\Delta^2$, the coefficients of states with $n \ge 1$ evolve much faster than the coefficients of states with $n=0$. This allows us to eliminate the cavity field adiabatically from the evolution of the system. Setting the derivative of the coefficients with $n = 1$ equal to zero, we find that
\begin{eqnarray} \label{eqn:coeffexcavity}
	c_{00;1} &=& 0 \, , \nonumber \\
	c_{01;1} &=& \frac{\Omega g}{2 \Delta^2 \kappa^2} \, \big( 2 {\rm i} \Delta \kappa - \Omega^2 - 4 g^2 - \kappa \Gamma \big) \, c_{01,0} \, , \nonumber\\
         c_{10;1} &=& \frac{\Omega g}{2 \Delta^2 \kappa^2} \big( 2 {\rm i} \Delta \kappa - \Omega^2 - 4 g^2 - \kappa \Gamma \big) \, c_{10,0} \, , \nonumber \\
	c_{11;1} &=& \frac{\Omega g}{\Delta^2 \kappa^2} \big( 2 {\rm i} \Delta \kappa - 2 \Omega^2 - 8 g^2 - \kappa \Gamma \big) \, c_{11,0} ~~
\end{eqnarray}
up to second order in $1/\Delta$. On average there is much less than one photon in the cavity mode.

Finally we derive a set of differential equations for the ground state coefficients of the atom-cavity system. Introducing the effective parameters 
\begin{eqnarray} \label{eqn:parity_effective_constants}
\Delta _{\rm eff} \equiv \frac{\Omega^2}{4\Delta } \, , ~~ \Gamma_{\rm eff} \equiv \frac{\Omega^2 \Gamma}{4\Delta^2} \, , 
~~ \kappa_{\rm eff} \equiv \frac{{\Omega^2 g^2}}{\Delta^2 \kappa } 
\end{eqnarray}
and substituting Eqs.~(\ref{eqn:coeffexatom}) and (\ref{eqn:coeffexcavity}) into Eq.~(\ref{eqn:coeffgeneral}), we obtain the effective differential equations
\begin{eqnarray}
\label{eqn:coeffsfinal}
	{\dot{c}}_{00;0} &= & 0 \, , \nonumber\\
	{\dot{c}}_{01;0} &=& \big( {\rm i} \Delta_{\rm eff} - {\textstyle \frac{1}{2}} \Gamma_{\rm eff} - {\textstyle \frac{1}{2}} \kappa_{\rm eff} \big) \, c_{01;0}  \, , \nonumber \\
	{\dot{c}}_{10;0} &=&  \big( {\rm i} \Delta_{\rm eff} - {\textstyle \frac{1}{2}} \Gamma_{\rm eff} - {\textstyle \frac{1 }{ 2}} \kappa_{\rm eff} \big) \, c_{10;0}  \, , \nonumber \\
	{\dot{c}}_{11;0} &=& \big( 2 {\rm i} \Delta_{\rm eff} - \Gamma_{\rm eff} - 2 \kappa_{\rm eff} \big) c_{11;0}  \, .
\end{eqnarray}
Here the spontaneous decay rates are correct up to second order in $1/\Delta$, while level shifts small compared to $\Delta_{\rm eff}$ have been neglected. Eq.~(\ref{eqn:coeffsfinal}) can be sumarised in the effective Hamiltonian 
\begin{eqnarray} \label{eqn:Hcondfinal}
H_{\rm eff} &=& - \hbar \big( \Delta _{\rm eff} + {\textstyle \frac{{\rm i} }{ 2}} \Gamma _{\rm eff} + {\textstyle \frac{{\rm i}}{2}} \kappa_{\rm eff} \big) \big[ |01\>\<01| + |10\>\<10| \big] \nonumber \\
&& - \hbar \big( 2 \Delta_{\rm eff} + {\rm i} \Gamma _{\rm eff} + 2 {\rm i} \kappa_{\rm eff} \big) \, |11\> \< 11| \, ,
\end{eqnarray} 
which acts only on the atomic ground states. The state $|00 \>$ is effectively decoupled from the dynamics of the system, while the states $|01\>$ and $|10\>$ evolve in the same way. Both cause an atomic emission with the effective decay rate $\Gamma_{\rm eff}$ or the leakage of a photon through the cavity mirrors with $\kappa_{\rm eff}$. The state $|11\>$ causes an atomic emission with the decay rate $2 \Gamma_{\rm eff}$ and the emission of a cavity photon with $4 \kappa_{\rm eff}$.

\subsection{The effect of a photon emission}

The effect of a photon emission on the state of the atom-cavity system can be described with the help of supplementary jump or {\em reset} operators $R_x$. If $|\psi \rangle$ is the state prior to a photon emission of type $x$, then $R_x |\psi \>$ is the (unnormalised) state immediately afterwards. For convenience we define the reset operators $R_x$ in the following such that 
\begin{eqnarray}
\label{eqn:parity_prob_density}
w_x(\psi) &=& \| \, R_x |\psi \> \, \|^2 \, ,
\end{eqnarray}
is the probability density for the corresponding emission to take place.

According to the quantum jump approach \cite{Hegerfeldt}, the reset operator for the emission of a photon via the 2--$j$ transition of the atoms is given by
\begin{eqnarray} \label{eqn:R0}
	R_{j} &=& \sqrt{\Gamma _j}\sum _{i=1,2}|j\> _{{ii}}\<2| \, , 
\end{eqnarray}
if the photons from the 2--0 and the 2--1 transition are distinguishable. The discussion in the previous subsection shows that the only states with population in the excited atomic state are $|02;0 \>$, $|20;0 \>$, $|12;0\>$, $|21;0 \>$, and $|22;0 \>$. From Eq.~(\ref{eqn:coeffexatom}) we see that that coefficients of these states depend to first order in $1/\Delta$ only on the coefficients $c_{00;0}$, $c_{01;0}$, $c_{10;0}$ and $c_{11;0}$. Combining Eqs.~(\ref{eqn:coeffexatom}) and (\ref{eqn:R0}), we see that the reset operators for the photon emission from the atoms are effectively and up to an overall phase factor given by
\begin{eqnarray} \label{eqn:R0eff}
R_{{\rm eff};0} &=& \sqrt{\Gamma_{{\rm eff};0}} \, \big[ |00\> \< 01| + |00\> \<10| + |01\> \< 11| \nonumber \\
&& \hspace*{1.3cm} + |10\> \< 11| \big] \, , \nonumber \\
R_{{\rm eff};1} &=& \sqrt{\Gamma _{{\rm eff};1}} \, \big[ |01\> \< 01| + |10\> \<10| + 2 \, |11\> \< 11| \big] ~~~~
\end{eqnarray}
with 
\begin{eqnarray}
\label{eqn:parity_gamma_eff}
\Gamma _{{\rm eff};j} &\equiv  & \frac{\Gamma_{j}\Gamma_{\rm eff}}{\Gamma } \, ,
\end{eqnarray}
and $\Gamma_{{\rm eff};0}+\Gamma_{{\rm eff};1} = \Gamma_{\rm eff}$. Again we find that the states $|01 \>$ and $|10 \>$ have the atomic decay rate $\Gamma_{\rm eff}$, while $|11 \>$ has the atomic decay rate $2 \Gamma_{\rm eff}$. 

During the leakage of a photon through the cavity mirrors with decay rate $\kappa$, one photon is removed from the resonator field. The corresponding reset operator is therfore given by
\begin{eqnarray} \label{eqn:Rcav}
R_{\rm C} &=& \sqrt{\kappa} \, b \, .
\end{eqnarray}
In the previous section we have seen that there is on average at most one photon in the cavity. Only the states $|01;1\>$,  $|01;1\>$, and $|11;1\>$ contribute to a cavity photon emission. We know that their coefficients depend only on $c_{01;0}$, $c_{10;0}$ and $c_{11;0}$. Combining Eqs.~(\ref{eqn:coeffexcavity}) and (\ref{eqn:Rcav}) we find that the cavity jump operator $R_{\rm C}$ is to first order in $1/\Delta$ given by
\begin{eqnarray} \label{eqn:Rcaveff}
R_{{\rm eff};C} &=& \sqrt{\kappa_{\rm eff}} \, \big[ |01\> \<01|+|10\> \< 10|+2 \, |11\> \< 11| \big] \, . ~~~~
\end{eqnarray}
Again, we see that the states $|01 \>$ and $|10 \>$ can cause a cavity photon emission with decay rate $\kappa_{\rm eff}$, while $|11 \>$ has the collectively enhanced cavity decay rate $4 \kappa_{\rm eff}$. 

\subsection{The master equation}

It should also be noted that the quantum jump approach above is consistent with the master equation, which is often alternatively used for the description of an open quantum system. It reads
\begin{eqnarray} \label{eqn:master}
\dot{\rho} = - {\textstyle \frac{\rm i}{\hbar}}\left[ H_{\rm cond} \rho - \rho H_{\rm cond }^{\dag}\right] + {\cal R}_{\rm eff}(\rho) \, ,
\end{eqnarray}
and can be obtained by averaging over all the possible trajectories that a single atom-cavity system can undergo \cite{Hegerfeldt}. The master equation is particularly well suited to the prediction of ensemble averages. For example, instead of displaying the existence of discrete levels in the fluorescence of a single system, the master equation can be used to predict the intensity of the emitted light averaged over all possible trajectories. 

\section{Basic principle} \label{sec:distinct}

\begin{figure}
\begin{minipage}{\columnwidth}
\begin{center}
\resizebox{\columnwidth}{!}{\rotatebox{0}{\includegraphics {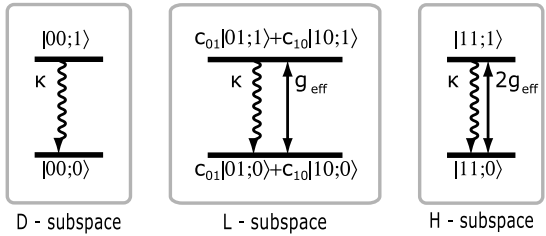}}}
\end{center}
\vspace*{-0.5cm}
\caption{Illustration of the effective evolution of the atoms. When spontaneous emission from the atoms is negligible, the state of the atoms converges within a relatively short time into one of the three subspaces $D$, $L$, and $H$. Each subspace is a characterised by a discrete fluorescence level in the leakage of photons through the cavity mirrors.} \label{fig:levels}
\end{minipage}
\end{figure}

Let us now discuss the dynamics of the atom-cavity system shown in Fig.~\ref{fig:setup} in more detail. Suppose the atoms are initially in
\begin{eqnarray} \label{eqn:initialstate_parity}
|\psi_0 \> &=& c_{00} \, |00 \> + c_{01} \, |01 \> + c_{10} \, |10 \> + c_{11} \, |11 \> \, . ~~~
\end{eqnarray}
As we see below, there are three distinct fluorescence levels in the leakage of photons through the cavity mirrors. Their origin is the existence of three decoupled subspaces in the effective evolution of the atomic ground states, when spontaneous emission from the atoms remains negligible. We denote them in the following by $D$, $L$ and $H$ (c.f.~Fig.~\ref{fig:levels}). The emission of photons at a certain rate or their complete absence gradually reveals information about the atoms. This gain of information gradually increases the population in one subspace with respect to the others until the population in any other subspace is irreversibly lost. The result is the projection of the atomic state $|\psi_0 \>$ into one subspace. 

\subsection{Three distinct fluorescence levels}

\begin{figure}
\begin{minipage}{\columnwidth}
\begin{center}
\resizebox{\columnwidth}{!}{\rotatebox{0}{\includegraphics{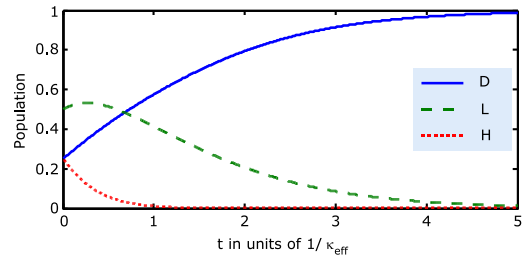}}}
\end{center}
\vspace*{-0.5cm}
\caption{(Colour online) A possible trajectory of the atom-cavity system given the initial state (\ref{eqn:parity_example1}) obtained from a quantum jump simulation with $\Gamma_0 = \Gamma_1 = 0.1 \, \kappa$, $g = \Omega = \kappa$, and $\Delta = 50\, \kappa$. Here no photons are emitted and all population eventually accumulates in the $D$-subspace.} \label{fig:parity00}
\end{minipage}
\end{figure}

Here the laser interaction is chosen such that the state $|00 \>$ is not involved in the evolution of the system. If no cavity photon is emitted for a time of the order of a few $1 / \kappa_{\rm eff}$, we therefore learn that the atoms are in a state where they cannot transfer population into the cavity mode. Consequently, the relative population in $|00 \rangle$ increases, while the population in the $L$ and in the $H$-subspace decrease. Eventually this results in the projection of the state (\ref{eqn:initialstate_parity}) into $|00 \>$ as shown in Fig.~\ref{fig:parity00}.  

\begin{figure}
\begin{minipage}{\columnwidth}
\begin{center}
\resizebox{\columnwidth}{!}{\rotatebox{0}{\includegraphics{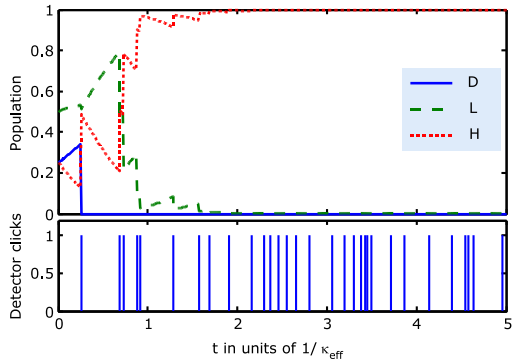}}}
\end{center}
\vspace*{-0.5cm}
\caption{(Colour online) A possible trajectory of the atom-cavity system for the same initial state and the same parameters as in Fig.~\ref{fig:parity00}. Here photons are emitted at a relatively high rate. This results eventually in a projection of the atomic state into the $H$-subspace.} \label{fig:parity11}
\end{minipage}
\end{figure}

\begin{figure}
\begin{minipage}{\columnwidth}
\begin{center}
\resizebox{\columnwidth}{!}{\rotatebox{0}{\includegraphics{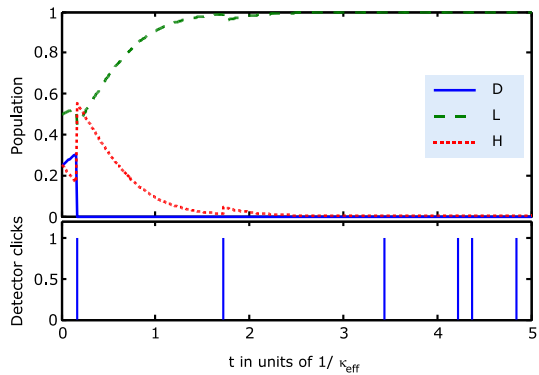}}}
\end{center}
\vspace*{-0.5cm}
\caption{(Colour online) A possible trajectory of the atom-cavity system for the same initial state and the same parameters as in Figs.~\ref{fig:parity00} and \ref{fig:parity11}. Here the time between two photon emissions is initially relatively long. This reveals that the atoms are most likely in the $L$-subspace and all population in $D$ and $H$ gradually vanishes.} \label{fig:parityLight}
\end{minipage}
\end{figure}

The occurrence of two distinct fluorescence periods when the atoms do lead to the emission of cavity photons may seem less obvious. Figs.~\ref{fig:parity11} and \ref{fig:parityLight} show possible trajectories of the system in these cases. Any population in $|00 \>$ vanishes with the first photon emission. Fig.~\ref{fig:parity11} shows how relatively frequent cavity emissions result in an ever increasing population in $|11 \>$, since this reveals that the atoms are most likely in the $H$-subspace. Eventually all atomic population accumulates in $|11 \>$, and can no longer return into another subspace via cavity photon emission. The system therefore continues to emit photons at its maximum rate given by $4 \kappa_{\rm eff}$.

However, it is also possible that the time between two subsequent photons is comparatively long. Such less frequent events result in a relative increase in population of the states $|01 \>$ and $|10 \>$ with respect to the population in $|11 \>$. This is illustrated in Fig.~\ref{fig:parityLight}. The reason for this is that seeing no photon for a time that is long compared to $1/4\kappa_{\rm eff}$ and after the first photon emission reveals that the system is more likely to be in the subspace with the lower emission rate. Eventually, the population in $|11 \>$ vanishes completely and the system emits cavity photons at the rate $\kappa_{\rm eff}$. Numerical simulations confirm that the probability of the odd-parity projection, $\proj_{01} + \proj_{10}$ (c.f.~Eq.~(\ref{eqn:parity})), indeed equals the initial population in the $L$-subspace, as predicted for an ideal measurement. 

\subsection{Implementation of a parity measurement} \label{sec:basicidea_parity}

In order to implement the parity measurement (\ref{eqn:parity}) in the absence of spontaneous emission from the atoms, it is sufficient to observe the cavity fluorescence over a time $T$ long enough to clearly distinguish the three fluorescence levels mentioned above. The concrete protocol is as follows:
\begin{enumerate}
\item  Place respective atoms inside the resonator.
\item Turn on driving laser for a time $T$ longer than a few $1 / \kappa_{\rm eff}$ and count the number of detected photons.
\end{enumerate}
The parity measurement is successful and results in the projection $\proj_{01} + \proj_{10}$, when the number of emitted photons is close to $\eta \kappa_{\rm eff} T$, which is the average number of photon detections when the system is in the $L$-subspace. The average number of photon detections when the system is in the $H$-subspace is given by $4 \eta \kappa_{\rm eff} T$. No photons indicate a projection onto the $D$-subspace. Fig.~\ref{fig:parity_simplefidprob} shows the fidelity and event probability for different detector click events for operation ({\ref{eqn:parity_example1}) and for a concrete choice of experimental parameters $(C=10)$. Even in the presence of a non-negligible spontaneous decay rate of the atoms, it is possible to achieve fidelities well above $0.9$.

\begin{figure}
\begin{minipage}{\columnwidth}
\begin{center}
\resizebox{\columnwidth}{!}{\rotatebox{0}{\includegraphics {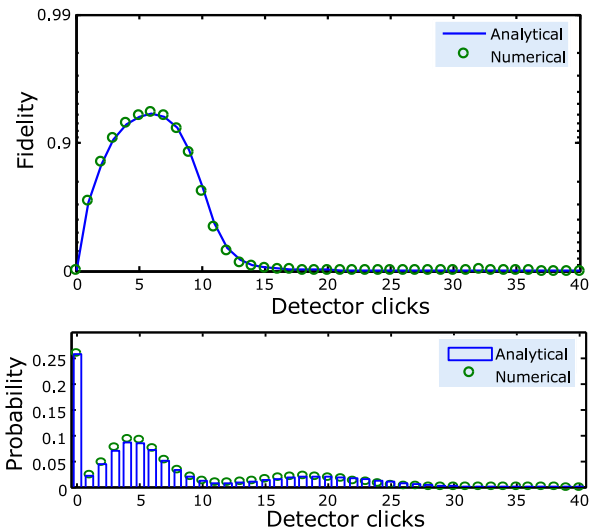}}}
\end{center}
\vspace*{-0.5cm}
\caption{(Colour online) Fidelity of the prepared state in case of $n$ detector clicks in $(0,T)$ averaged over many trajectories and the corresponding probability for this to happen. Here we compare the results from an analytical calculation \cite{Jeremy} with the results obtained from a quantum trajectory simulation with $T = 5 / \kappa_{\rm eff}$, $\Gamma_0 = \Gamma_1 = 0.05 \kappa$, $g=\Omega = \kappa$, $\Delta = 50 \kappa$, and  $\eta=1$. A photon number close the $\kappa_{\rm eff} T$ indicates that the atoms are most likely in the $L$ subspace, while no detector clicks or a relatively large number of clicks shows that the atoms are in $|00 \>$ or $|11 \>$, respectively.} \label{fig:parity_simplefidprob}
\end{minipage}
\end{figure}

\subsection{Optimised protocol} \label{sec:bitflip}

Deviations in the fidelity from unity in Fig.~\ref{fig:parity_simplefidprob} are largely due to the possibility of photon emission from the atoms with $\Gamma_{\rm eff}$. For example, when starting in $|11 \rangle$, the emission of a photon with reset operator $R_{{\rm eff};0}$ in Eq.~(\ref{eqn:R0eff}) projects the atoms onto the $L$-subspace. Similarly, such an emission can transfer population from the $L$-subspace into the $D$-subspace. In both cases, it might be assumed that the atoms are in $L$, consequently resulting in a decrease of the fidelity of the prepared state. When using the parity measurement for the generation of multi-qubit entanglement, as described for example in Eq.~(\ref{eqn:parity_example2}), the result is the loss of the entanglement with atoms outside the cavity.   

Reducing the occurrence of atomic emissions requires shortening the interaction time $T$. However, in the above protocol, this would make it difficult to distinguish the fluorescence when in $H$ from that in $L$. As a solution we propose to use the double heralding technique of Barrett and Kok \cite{Kok}. We now consider the following variant of the above protocol (c.f.~Fig.~\ref{fig:parity_opt_protocol}):
\begin{enumerate}
\item Place respective atoms inside the resonator.
\item Turn on driving laser for a maximum time $T_{\rm max}$ or until the first detection of a photon. 
\item Swap the states $|0\>$ and $|1 \>$ of each atom.
\item Repeat step 2 and 3.
\end{enumerate}
This protocol allows us to measure the parity of atoms in a relatively short time. In the ideal case, the only events that produce two detector clicks are due to the system being in $L$. The reason for this is that atoms in $|00 \>$ cannot produce a click in step 2. Atoms initially in $|11 \rangle$ are transferred into $|00 \>$ in step 3 and are therefore unable to emit a photon in step 4.

\begin{figure}
\begin{minipage}{\columnwidth}
\begin{center}
\resizebox{\columnwidth}{!}{\rotatebox{0}{\includegraphics {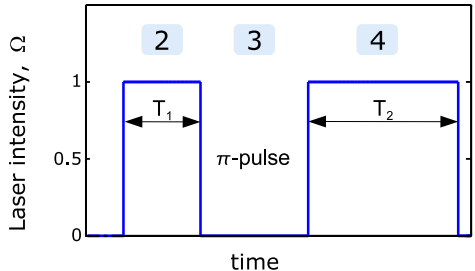}}}
\end{center}
\vspace*{-1cm}
\caption{(Colour online) Schematic view of the main stages of the optimised protocol. After the atoms have been placed into the resonator in step 1, the laser field is switched on in step 2 until a cavity photon is detected at time $T_1$. In step 3 a $\pi$-pulse transfers $|0\>$ into $|1\>$ and vice-versa. In step 4 the laser is turned on again until a second click at $T_2$. This second click signals an odd parity state of the atoms.} \label{fig:parity_opt_protocol}
\end{minipage}
\end{figure}

Finally we comment on the optimal size of $T_{\rm max}$. The duration of the applied laser pulse should be comparable to the mean time between two photon detections when the system is in the $L$-subspace. This is sufficient to assure that there is a high probability for the detection of a photon in step 2 and 4, if the atoms are in $L$. For example, if $T_{\rm max} = 3/\eta \kappa_{\rm eff}$,  this probability is already above $90 \, \%$. Longer laser pulses do not increase this probability significantly and can therefore be avoided. They might only lead to a slight decrease of the fidelity of the odd-parity measurement due to an increase in the probability for the spontaneous emission of a photon from one of the atoms before the detection of the first cavity photons. In the remainder of this paper, we nevertheless assume that $T_{\rm max} = \infty$. This simplifies the following calculations which nevertheless yield good approximations for the actual fidelity and the success rate of the proposed scheme for finite $T_{\rm max}$.

\section{Performance analysis of the optimised protocol}\label{sec:parity}

In the following we analyse the optimised protocol in detail and show that its performance is comparable to the scheme presented in Ref.~\cite{MTB-PRL}.  As an example, we consider the operation described in Eq.~(\ref{eqn:parity_example2}) and calculate the average fidelity and success rate for the preparation of a four-atom GHZ state. In Section \ref{sec:FT1T2} we analyse the ideal scenario of perfect photon detection. Finite photon detector efficiencies are taken into account in Section \ref{sec:parity_eta}. In Section \ref{sec:parity_robust} we emphasize that the performance of the proposed state preparation scheme is essentially independent of the concrete size of the system parameters and hence very robust against parameter fluctuations.

\subsection{Average fidelity for unit efficiency photon detectors} \label{sec:FT1T2}

One factor that decreases the fidelity of the prepared state is population in excited atomic states after the laser field has been turned off. This population might result in an atomic emission, which transfers the system into a state with a reduced overlap with the target state. From Eq.~(\ref{eqn:coeffexatom}) we see that it equals
\begin{eqnarray} \label{eqn:PopEx} 
P_{\rm atom~excited} &=& \frac{\Omega^2 }{ 4 \Delta^2} \, ,
\end{eqnarray}
up to second order in $1/\Delta$, in the case of the odd parity projection $\proj_{01} + \proj_{10}$. This population can be made arbitrarily small even in the presence of relatively large spontaneous decay rates by simply increasing the detuning $\Delta$. In the parameter regime (\ref{eqn:conditions}), corrections due to the population in Eq.~(\ref{eqn:PopEx}) are hence negligible. Possible remaining cavity excitations do not affect the fidelity of operation (\ref{eqn:parity_example2}), since their emission does not affect the state of the atoms once the projection onto one of the subspaces $D$, $L$ or $H$ has occurred.

Corrections to the fidelity of the state prepared through an odd-parity projection are generally dominated by effects due to spontaneous emission from the atoms. To calculate these corrections for unit efficiency photon detectors, we consider the following two events:
\begin{itemize} 
\item {\em Event} $A$: The atoms {\em emit} the first cavity photon at $T_1$ in step 2 and at $T_2$ in step 4. Spontaneous emissions from excited atomic states may occur but the atoms are finally in the desired state.
\item {\em Event} $B$: The atoms {\em emit} the first cavity photon at $T_1$ in step 2 and at $T_2$ in step 4. Spontaneous emissions from excited atomic states occur and the final state of the system is {\em not} the desired one.
\end{itemize}
If $w_X(T_1,T_2)$ denotes the probability density for having event $X$, the fidelity of a successful odd parity check equals 
\begin{eqnarray} \label{eqn:fopt_def}
F(T_1, T_2) &=& \frac{w_A(T_1,T_2)}{w_A(T_1,T_2) + w_B(T_1,T_2)} \, .
\end{eqnarray}
We now analyse this fidelity in terms of the probabilities of Markovian processes.

To do so, we denote the probability density for the spontaneous emission of a photon from the atoms, which transfers the atoms from the $X$-subspace into the $Y$-subspace, by $\gamma_{XY}$. Moreover, $\kappa_X$ is the probability density for a cavity photon emission, when the atoms are in $X$. From Eq.~(\ref{eqn:Rcaveff}) we see that such an emission does not change the state of the atoms within the respective subspace. Given an initial state, where the coefficients of the $|00 \>$, $|01 \>$, $|10 \>$ and $|11 \>$ components of the state of the atoms inside the cavity are initially all the same (as in Eq.~(\ref{eqn:parity_example2})), the probability density $w_A(T_1,T_2)$ can be written as
\begin{eqnarray} \label{eqn:wA_def}
w_A(T_1, T_2) &=& {\textstyle \frac{1}{2}} \sum_{n=0}^\infty \kappa_L \, {\rm e}^{- \kappa_L T_1} \, P_n^{(L)}(T_1) \nonumber \\
&& \times \sum_{m=0}^\infty \kappa_L \, {\rm e}^{- \kappa_L T_2} \, P_m^{(L)}(T_2) \, F(n+m) \nonumber \\
&=& {\textstyle \frac{1}{2}}  \sum_{n=0}^\infty \kappa_L^2 \, {\rm e}^{- \kappa_L (T_1+T_2)} \, P_n^{(L)}(T_1+T_2) F(n) \, . \nonumber \\
\end{eqnarray}
Here the factor ${\textstyle \frac{1}{2}}$ is the initial population in the $L$-subspace and ${\rm e}^{- \kappa_L T}$ is the probability for no cavity decay in $(0,T)$ given that the atoms are in $L$. Moreover, $P_n^{(L)}(T)$ is the probability of $n$ atomic emissions from subspace $L$ back into $L$ and no emissions from $L$ into $D$ in $(0,T)$. An expression for this probability can be found in Eq.~(\ref{eqn:parity_PnL_app2}). The fidelity of the final state after $n$ such spontaneous emissions is denoted by $F(n)$. Here $F(0)=1$ and $F(n>0) = {\textstyle \frac{1}{2}}$ due to the nature of spontaneous emission from the atoms.

Similarly we now calculate the probability density for event $B$ to occur. Doing so we find that
\begin{eqnarray} \label{eqn:wB_def}
w_B(T_1, T_2) &=& {\textstyle \frac{1}{4}} \int_0^{T_1} {\rm d}t \sum_{m=0}^\infty \gamma_{HL} \, {\rm e}^{-\kappa_H t} P_m^{(H)}(t) \nonumber \\
&& \hspace*{-1.2cm} \times \sum_{n=0}^\infty \kappa_L^2 \, {\rm e}^{- \kappa_L (T_1 - t + T_2)} \, P_n^{(L)}(T_1 - t + T_2) \, . \nonumber \\
&& \hspace*{-1.2cm}  +  {\textstyle \frac{1}{2}} \sum_{n=0}^\infty \kappa_L^2 \, {\rm e}^{- \kappa_L (T_1 + T_2)} \, P_n^{(L)}(T_1 + T_2) (1- F(n)) \, . \nonumber \\
\end{eqnarray}
The factor ${\textstyle \frac{1 }{4}}$ in the first line is the initial population in the $H$-subspace. Here $t$ is the time at which a transition from $H$ to $L$ occurs. To calculate $w_B(T_1, T_2)$, we integrate over all possible values for $t$. Moreover, ${\rm e}^{- (\gamma_{HL} + \kappa_H) t}$ is the probability for no atomic emission out of $H$ and no cavity decay in $(0,t)$ given that the atoms are in $H$. In analogy to the above notation, $P_n^{(H)}(t)$ is the probability of $n$ atomic emissions from subspace $H$ back into $H$ and no emissions from $H$ into $L$ in $(0,t)$. An expression for this probability can be found in Eq.~(\ref{eqn:parity_PnH_app}). The last line in Eq.~(\ref{eqn:wB_def}) is analoguos to Eq.~(\ref{eqn:wA_def}) and contains the probability density that the final state of the atoms does {\em not} overlap with the desired state although the atoms were initially in $L$ in the absence of a transition from the $H$ into the $L$-subspace. 

Using Eqs.~(\ref{eqn:parity_PnL_app1}) and (\ref{eqn:parity_PnH_app}), the probability densities $w_A(T_1, T_2)$ and $w_B(T_1, T_2)$ can be calculated analytically. Doing so, we obtain
\begin{eqnarray} \label{eqn:wA_app}
w_A(T_1, T_2) &=& {\textstyle \frac{1}{ 4}} \kappa_L^2 \Big[ {\rm e}^{- ( \kappa_L + \gamma_{LD} + \gamma_{LL}) (T_1 + T_2)} \nonumber \\
&& + {\rm e}^{- (\kappa_L + \gamma_{LD}) (T_1 + T_2)} \Big] \, ,
\end{eqnarray}
and 
\begin{eqnarray} \label{eqn:wB_app}
w_B(T_1, T_2) &=& {\textstyle \frac{1}{4}} \kappa_{L}^2 \Bigg[ \frac{\gamma_{HL}} {\kappa_H - \kappa_L + \gamma_{HL} - \gamma_{LD}} \nonumber \\
&& \hspace*{-1.9cm} \times \Big( {\rm e}^{-(\kappa_L + \gamma_{LD})(T_1 + T_2)} - {\rm e}^{-(\kappa_L + \gamma_{LD})T_2} \, {\rm e}^{- (\kappa_H+\gamma_{HL}) T_1} \Big) \nonumber \\
&& \hspace*{-1.9cm} - {\rm e}^{-(\kappa_L + \gamma_{LD}+\gamma_{LL})(T_1+T_2)} + {\rm e}^{- (\kappa_L  + \gamma_{LD})(T_1+T_2)} \Bigg] \, . 
\end{eqnarray}
It is possible to optimise the corresponding fidelity $F(T_1,T_2)$ in Eq.~(\ref{eqn:fopt_def}) by postselecting events, where $T_1$ and $T_2$ are both short. In such cases, the probability for a decrease of the fidelity due to an atomic emission is low and $w_B(T_1, T_2)$ remains negligible. However, this optimisation comes at the cost of a decrease of the success rate for an odd parity check. 

To maximise the efficiency of the proposed cluster state preparation scheme, let us accept all events independent of the size $T_1$ and $T_2$ and assume that the interaction time $T_{\rm max}$ in Fig.~\ref{fig:parity_opt_protocol} is very large. The average fidelity of the final state in the case of a projection onto the $L$-subspace is then given by
\begin{eqnarray} \label{eqn:parity_Fav}
F_{\rm av} &=& \frac{P_A}{P_A + P_B} \, ,
\end{eqnarray}
in analogy to Eq.~(\ref{eqn:fopt_def}), if $P_X$ is the probability for an event $X=A,B$ to take place. Moreover 
\begin{eqnarray} \label{eqn:parity_ProbSucc}
P_{\rm suc} &=& P_A + P_B 
\end{eqnarray}
is the probability for the observation of such a projection and equals the denominator in Eq.~(\ref{eqn:parity_Fav}). Since
\begin{eqnarray} \label{eqn:parity_ProbX}
P_X &=& \lim_{T_{\rm max} \to \infty} \int_0^{T_{\rm max}} {\rm d}T_1 \int_0^{T_{\rm max}} {\rm d}T_2 \,  w_X(T_1, T_2) \, , ~~~~~
\end{eqnarray}
we find, using Eqs.~(\ref{eqn:wA_app}) and (\ref{eqn:wB_app}), 
\begin{eqnarray} \label{eqn:parity_ProbA}
P_A &=& {\textstyle \frac{1}{4}} \kappa_{L}^2 \left[ \frac{1} {(\kappa_L + \gamma_{LD} + \gamma_{LL})^2} + \frac{1} {(\kappa_L + \gamma_{LD})^2} \right] \, , \nonumber \\
P_B &=& {\textstyle \frac{1}{4}} \kappa_{L}^2 \, \frac{1}{(\kappa_L + \gamma_{LD})^2} \left[ \frac{\gamma_{LL}(\gamma_{LL} + 2 \kappa_L + 2 \gamma_{LD})}{(\kappa_L + \gamma_{LD}+\gamma_{LL})^2} \right. \nonumber \\
&& \left. + \frac{\gamma_{HL}}{\kappa_H + \gamma_{HL}} \right]  \, ,
\end{eqnarray}
with the transition rates (c.f.~Section \ref{sec:theory})
\begin{eqnarray} \label{eqn:parity_gammas_def}
\gamma_{HL} = 2 \gamma_{LL} &=& 2 \Gamma_{\rm eff;1} \, , \nonumber \\
\gamma_{HH} = 2 \gamma_{LD} &=& 2 \Gamma_{\rm eff;0} \, , \nonumber \\
\kappa_H = 4 \kappa_L &=& 4 \kappa_{\rm eff} \, . 
\end{eqnarray}
In the special case, where $\Gamma_0 = \Gamma_1$, these equations simplify to
\begin{eqnarray} \label{eqn:parity_gammas_relations}
2 \gamma_{LL} = 2 \gamma_{LD} = \gamma_{HL} &=& \gamma_{HH}  \, , \nonumber \\
\kappa_H = 4 \kappa_L &=& 16 C \, \gamma_{HH} \, ,
\end{eqnarray}
and the average fidelity and the success rate depend only on the single atom cooperativity parameter $C$,
\begin{eqnarray} \label{eqn:Fav_final}
F_{\rm av} &=& 
	\frac{5/32 + 4 C + 28 C^2 + 64 C^3}
	{3/8 + 7 C + 38 C^2 + 64 C^3}  
\end{eqnarray}
and
\begin{eqnarray} \label{eqn:Psucc_final}
P_{\rm suc} &=& 
	\frac{6 C^2 + 64 C^3}
	{1/8 + 4 C + 40 C^2 + 128 C^3} \, .
\end{eqnarray}
In the limit of large $C$, the fidelity $F_{\rm av}$ of the entangling operation (\ref{eqn:parity_example2}) approaches unity, while the success rate $P_{\rm suc}$ converges to ${\textstyle \frac{1}{2}}$, as expected for an ideal projection. In the presence of non-negligible atomic emission, i.e.~for smaller $C$'s, the fidelity and success rate are slightly smaller. Even when $C=1$, we have $F_{\rm av} = 0.88$ and $P_{\rm suc} = 0.41$ in the presence of ideal photon detectors with $\eta=1$. Average fidelities exceeding 0.99 become possible when $C$ approaches $20$.
 
\subsection{Average fidelity for finite efficiency photon detectors} \label{sec:parity_eta}

\begin{figure}
\begin{minipage}{\columnwidth}
\begin{center}
\resizebox{\columnwidth}{!}{\rotatebox{0}{\includegraphics {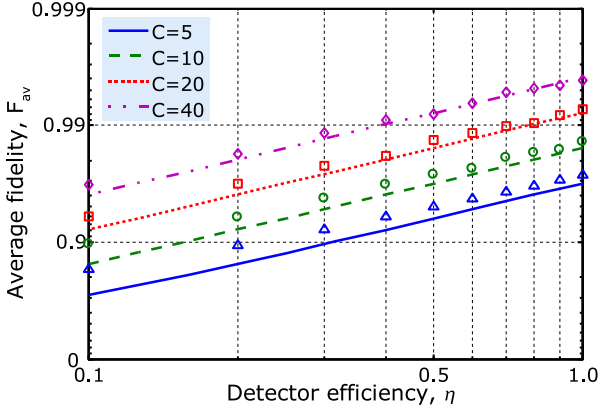}}}
\end{center}
\vspace*{-0.5cm}
\caption{(Colour online) Logarithmic plot of the fidelity of the output state against the detector efficiency $\eta$ for several values of the cooperativity parameter $C$. The lines show the analytical solution given in Eq. (\ref{eqn:Fav_eta}) for the four-atom GHZ state preparation described in Eq.~(\ref{eqn:parity_example2}), while the symbols have been obtained from a quantum trajectory simulation of the simpler entangling operation (\ref{eqn:parity_example1}) assuming $\Delta = 50 \, \kappa$, and $g = \Omega = \kappa$. The triangle, circle , square, and diamond, correspond to the $C=5$, $10$, $20$, and $40$, respectively.} \label{fig:parity_fidnumerics}
\end{minipage}
\end{figure}

To calculate the average fidelity and success rate for an odd-parity projection for finite photon detector efficiencies $\eta < 1$, we consider the events:
\begin{itemize} 
\item {\em Event} $A'$: The first {\em detection} of a cavity photon at $T_1$ in step 2 and at $T_2$ in step 4. Spontaneous emissions from excited atomic states may occur but the atoms are finally in the desired state.
\item {\em Event} $B'$: The first {\em detection} of a cavity photon at $T_1$ in step 2 and at $T_2$ in step 4. Spontaneous emissions from excited atomic states occur and the final state of the system is {\em not} the desired one.
\end{itemize}
We then notice that $\eta $ has no effect on the probability densities for atomic emissions. It only extends the mean time until the detection of a first cavity photons. To obtain the probability densities $w_{A'} (T_1,T_2)$ and $w_{B'} (T_1,T_2)$, we therefore only need to replace $\kappa_X$  in Eqs.~(\ref{eqn:wA_def}) and (\ref{eqn:wB_def}) by $\eta \kappa_X$ and ${\rm e}^{- \kappa_X t}$ by ${\rm e}^{- \eta \kappa_X t}$. Proceeding as above, we then find that the average fidelity and success rate are now given by 
\begin{eqnarray} \label{eqn:Fav_eta}
F_{\rm av} &=& 
	\frac{5/32 + 4 \eta C + 28 (\eta C)^2 + 64 (\eta C)^3}
	{3/8 + 7 \eta C + 38 (\eta C)^2 + 64 (\eta C)^3}
\end{eqnarray}
and
\begin{eqnarray} \label{eqn:Psucc_eta}
P_{\rm suc} &=& 
	\frac{6 (\eta C)^2 + 64 (\eta C)^3}
	{1/8 + 4 \eta C + 40 (\eta C)^2 + 128 (\eta C)^3} \, .
\end{eqnarray}
These are the same expressions as in Eqs.~(\ref{eqn:Fav_final}) and (\ref{eqn:Psucc_final}) but with $C$ now replaced by $\eta C$.

\begin{figure}
\begin{minipage}{\columnwidth}
\begin{center}
\resizebox{\columnwidth}{!}{\rotatebox{0}{\includegraphics {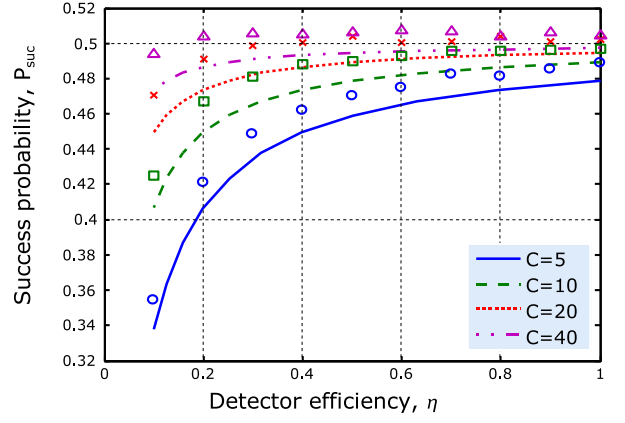}}}
\end{center}
\vspace*{-0.5cm}
\caption{(Colour online) Probability of a successful odd parity check as a function of $\eta$, for several values of $C$ and for the same parameters as in Fig.~\ref{fig:parity_fidnumerics}. The circle , square, cross, and triangle, correspond to the $C=5$, $10$, $20$, and $40$, respectively. The lines show the analytical solution given in Eq. (\ref{eqn:Psucc_eta}) for the four-atom GHZ state preparation described in Eq.~(\ref{eqn:parity_example2}).} \label{fig:parity_prob_opt}
\end{minipage}
\end{figure}

Figs.~\ref{fig:parity_fidnumerics} and \ref{fig:parity_prob_opt} compare the average fidelity in Eq.~(\ref{eqn:Fav_eta}) and the success rate for an odd-parity projection in Eq.~(\ref{eqn:Psucc_eta}) with the results obtained from a numerical solution of the evolution of the system using Eqs.~(\ref{eqn:Hcond}), (\ref{eqn:R0}), and (\ref{eqn:Rcav}) for the entangling operation described in Eq.~(\ref{eqn:parity_example1}). Operation (\ref{eqn:parity_example1}) is slightly different from operation (\ref{eqn:parity_example2}), which we considered in the above calculations, since atomic emissions lead less frequently to an error in the simpler pair entangling scheme, as there is no initial entanglement that needs to be preserved. Nevertheless, there is relatively good agreement between both curves. Figs.~\ref{fig:parity_fidnumerics} and \ref{fig:parity_prob_opt} indeed confirm that reducing $\eta$ has the same effect as replacing $C$ by $\eta C$. 

\subsection{Parameter dependence} \label{sec:parity_robust}

Due to its postselective nature, the performance of the proposed state preparation scheme is essentially independent of the concrete system parameters. Fig.~\ref{fig:parity_fidnumerics} shows that fidelities well above $99 \, \%$ are inevitably when $\eta C > 20$. Whenever this condition is fulfilled, there are three distinct fluorescence levels in the emission of cavity photons. The scheme is constructed such that the medium level always indicates that one atom is in $|0 \rangle$ and one atom is in $|1 \rangle$ without revealing which one. Turning off the applied laser field upon the detection of a photon in step 2 and 4 is hence sufficient to realise the parity operation (\ref{eqn:parity}) with very high accuracy. The proposed state preparation scheme is therefore robust against a parameter fluctuations, like moderate fluctuations of cavity coupling constants and laser Rabi frequencies.  

In the following, we show that high fidelities are achieved even when the atoms experience quite different coupling constants. As an example, we consider the case where the cavity-coupling constant of atom 1 is $g_1$ and the cavity-coupling constant of atoms 2 is given by $g_2$. For simplicity we neglect spontaneous emission from the atoms $(\Gamma = 0)$ in the following. Proceeding as in Section \ref{sec:theory}, we find that the conditional Hamiltonian (\ref{eqn:Hcondfinal}) is now given by 
\begin{eqnarray} \label{eqn:Heff_robust}
H_{\rm eff} &=& - \hbar \big( \Delta _{\rm eff} + {\textstyle \frac{{\rm i}}{2}} \kappa_{\rm eff;2} \big) \, |01\>\<01| \nonumber \\
&& - \hbar \big( \Delta _{\rm eff} + {\textstyle \frac{{\rm i}}{2}} \kappa_{\rm eff;1} \big) \, |10\>\<10| \nonumber \\
&& - \hbar \big[ 2 \Delta_{\rm eff} + {\textstyle \frac{\rm i}{2}} 
(\sqrt{\kappa_{\rm eff;1}} + \sqrt{\kappa_{\rm eff;2}} \, )^2 \big] \, |11\> \< 11| ~~~~~~
\end{eqnarray} 
with
\begin{eqnarray} \label{eqn:eff_const_robust}
\kappa_{{\rm eff};i} &\equiv & \frac{\Omega^2 g_{i}^2}{\Delta^2 \kappa} \, .
\end{eqnarray} 
At the same time, the reset operator in Eq.~(\ref{eqn:Rcaveff}) becomes
\begin{eqnarray} \label{eqn:Reset_robust}
R_{{\rm eff};C} &=& \sqrt{ \kappa_{\rm eff;1}} \, \big( |10\> \< 10| +  |11\> \< 11| \big) \nonumber \\
&& + \sqrt{\kappa_{\rm eff;2}} \, \big( |01\> \<01| + |11\> \< 11| \big) \, . 
\end{eqnarray} 
These two equations can be used to simulate for example all the possible trajectories for the two-qubit entangling operation (\ref{eqn:parity_example1}).

\begin{figure}
\begin{minipage}{\columnwidth}
\begin{center}
\resizebox{\columnwidth}{!}{\rotatebox{0}{\includegraphics {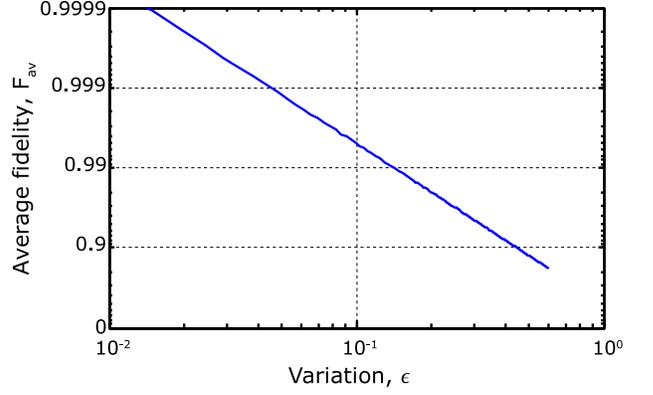}}}
\end{center}
\vspace*{-0.5cm}
\caption{(Colour online) Average fidelity, $F_{\rm av}$, from Eq.~(\ref{eqn:parity_Fav_robust_final}) for the entangling operation described in Eq.~(\ref{eqn:parity_example1}) as a function of $\epsilon$. Fidelities above 0.9 are achievable, even when $\kappa_{\rm eff;1}$ and $\kappa_{\rm eff;2}$ differ from each other by $\bar \kappa_{\rm eff}$.} \label{fig:parity_fidrobust}
\end{minipage}
\end{figure}

Let us first consider the case, where the atoms {\em permanently} see different coupling constants $g_i$. Fig.~\ref{fig:parity_fidrobust} shows the average fidelity $F_{\rm av}$ of the prepared state in the case of an odd parity projection as a function of $\epsilon$, where
\begin{eqnarray} \label{eqn:parity_robust_kappaeff}
&& \kappa_{\rm eff;1} \equiv (1 + \epsilon) \, \bar \kappa_{\rm eff} \, , ~~
\kappa_{\rm eff;2} \equiv (1 - \epsilon) \, \bar \kappa_{\rm eff} \, , ~~ \nonumber \\
&& \bar \kappa_{\rm eff} \equiv {\textstyle \frac{1}{2}} \big( \kappa_{\rm eff;1} + \kappa_{\rm eff;2} \big) \, .
\end{eqnarray}
Fidelities above 0.9 are achievable, even when $\kappa_{\rm eff;1}$ and $\kappa_{\rm eff;2}$ differ by their average value ${\bar \kappa}_{\rm eff}$! The reason for this is the postselective nature of the proposed optimised protocol.

To illustrate this we now examine the dynamics of the system in the absence of spontaneous emissions from the atoms in more detail. The state of the atoms in case of a detector click at $T_1$ and one at $T_2$ is given by
\begin{eqnarray} \label{eqn:parity_robust_state}
&& \hspace*{-0.7cm} |\psi(T_1,T_2) \> \nonumber \\
&=& \frac{R_{\rm eff;C} \, U_{\rm cond} (T_2,0) \, U_\pi \, R_{\rm eff;C} \, U_{\rm cond} (T_1,0) \, |\psi_0 \>}{\| \, ... \,  \|} ~~~~
\end{eqnarray}
with $U_\pi$ being the operation that exchanges the states $|0 \>$ and $|1 \>$. Using Eqs.~(\ref{eqn:Heff_robust}) and (\ref{eqn:Reset_robust}), we find that the system is finally in the state
\begin{eqnarray} \label{eqn:parity_robust_state2}
|\psi(T_1,T_2)\> 
&=& \frac{1}{\| \, ... \, \|} \left( {\rm e}^{- (\kappa_{\rm eff;2} T_1 + \kappa_{\rm eff;1}T_2)/2} \, |01\> \right. \nonumber \\
&& \left. + {\rm e}^{-(\kappa_{\rm eff;1} T_1 + \kappa_{\rm eff;2}T_2)/2} \, |10\>  \right) \, .
\end{eqnarray}
Calculating the overlap of this state with the Bell state in Eq.~(\ref{eqn:parity_example1}), we obtain the fidelity 
\begin{eqnarray} \label{eqn:parity_Ft1t2_robust}
	&& \hspace*{-0.7cm} F(T_1,T_2) \nonumber \\
	&=& {\textstyle \frac{1}{2}} + \frac{ {\rm e}^{-(\kappa_{\rm eff;1}+\kappa_{\rm eff;2})(T_1+T_2)/2}}
	{{\rm e}^{-(\kappa_{\rm eff;1}T_1 + \kappa_{\rm eff;2}T_2)} 
	+ {\rm e}^{- (\kappa_{\rm eff;2}T_1 + \kappa_{\rm eff;1}T_2)}} \, . ~~~~~
\end{eqnarray}
To calculate the average fidelity for the entangling operation (\ref{eqn:parity_example2}), we notice that the probability density for a click at $T_1$ and $T_2$ is in this case given by
\begin{eqnarray}
\label{eqn:parity_wt1t2_robust}
w(T_1,T_2) &=& {\textstyle \frac{1}{4}} \, 	\kappa_{\rm eff;1} \kappa_{\rm eff;2} \, \left( \, {\rm e}^{- (\kappa_{\rm eff;2} T_1 + \kappa_{\rm eff;1} T_2) } \right. \nonumber \\
 && \left. + {\rm e}^{ - (\kappa_{\rm eff;1} T_1 + \kappa_{\rm eff;2} T_2)} \, \right) \, . ~~~
\end{eqnarray} 
The average fidelity $F_{\rm av}$ is again obtained by integrating over all possible click times $T_1$ and $T_2$,  
\begin{eqnarray}
	\label{eqn:parity_Fav_robust_def}
	F_{\rm av} &=&
	\frac{\int_0^{\infty} dT_1 \int_0^{\infty} dT_2 \, w(T_1,T_2) \, F(T_1,T_2)}
	{\int_0^{\infty} dT_1 \int_0^{\infty} dT_2 \, w(T_1,T_2)} \, .
\end{eqnarray}
Inserting Eqs.~(\ref{eqn:parity_robust_kappaeff}), (\ref{eqn:parity_Ft1t2_robust}) and (\ref{eqn:parity_wt1t2_robust}) into this equation, we find that 
\begin{eqnarray} \label{eqn:parity_Fav_robust_final}
F_{\rm av} &=& 1 - {\textstyle \frac{1}{2}} \epsilon^2\, .
\end{eqnarray}
This means that very large fidelities are possible even for non-negligible $\epsilon$ (c.f.~Fig.~\ref{fig:parity_fidrobust}).

The very high fidelities in Fig.~\ref{fig:parity_fidrobust} are due to the concrete form of the state of the atoms after two photon emissions in Eq.~(\ref{eqn:parity_robust_state2}). As one can see, the states $|01 \>$ and $|10 \>$ have approximately the same coefficients, when $\kappa_{\rm eff;2} T_1 + \kappa_{\rm eff;1} T_2$ and $\kappa_{\rm eff;1} T_1 + \kappa_{\rm eff;2} T_2$ are of about the same size. This applies for a very wide range of click times $T_1$ and $T_2$ and decay rates $\kappa_{\rm eff;1}$
and $\kappa_{\rm eff;2}$. Proceeding analogously, one can show that the proposed realisation of the odd-parity check (\ref{eqn:parity}) is robust against {\em temporal} fluctuations of the atom-cavity coupling constant $g$ and the laser Rabi frequency $\Omega$. 

\section{Cluster state growth}\label{sec:cluster}

In this section, we describe how to use the probabilistic parity check in Eq.~(\ref{eqn:parity}) for the build up of two-dimensional cluster states \cite{cluster}. These highly entangled states constitute the main resource for one-way quantum computing \cite{oneway,cluster}. Once a cluster state has been built, local operations and single-qubit read out measurements are sufficient to realise any possible quantum algorithm. That the projection (\ref{eqn:parity}) can be used to build cluster states has already been noted by Browne and Rudolph \cite{Rudolph}. 
Below we proceed in a similar fashion.

\subsection{Fusion of one-dimensional clusters} \label{fusion}

\begin{figure}
\begin{minipage}{\columnwidth}
\begin{center}
\resizebox{\columnwidth}{!}{\rotatebox{0}{\includegraphics{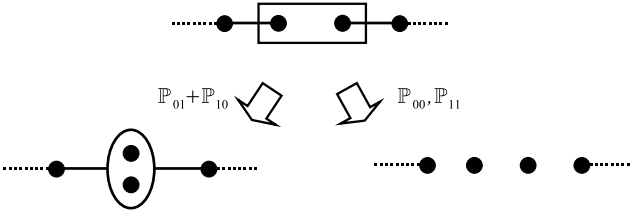}}}
\end{center}
\vspace*{-0.5cm} \caption{Fusion of two linear cluster states. A successful parity check creates a double-encoded qubit which links the two chains. If the parity check fails we project either on $|00\>$ or on $|11\>$. In both cases one qubit from each chain decouples.} \label{fig:1d_fusion}
\end{minipage}
\end{figure}

Let us first have a closer look at the build up of a linear cluster state, which consists of a chain of atoms with next neighbor entanglement. Larger clusters can be obtained through the {\em fusion} of two cluster states \cite{Rudolph}. In the scheme proposed here this requires placing one end atom from each chain into the optical cavity and performing the parity measurement (\ref{eqn:parity}). Entanglement between these two atoms is obtained {\em and} the original correlations with the atoms outside the cavity are preserved in case of a projection with $\proj_{01}+ \proj_{10}$. In case of a projection with $\proj_{00}$ or $\proj_{11}$, the atoms in the cavity decouple from the rest and a new attempt has to be made to incorporate them into a larger cluster.

More concretely, we now consider the case of fusing two cluster chains with $n-m$ and $m$ qubits, respectively, into one linear cluster of size $n-1$. The initial state of the atoms can then be written as \cite{cluster}
\begin{eqnarray} 
|\psi \> &=& \frac{1}{2^{n/2}} \bigotimes_{i=m+1}^{n} \big(|0\>_i+ \sigma_z^{(i-1)} |1\>_i \big) \nonumber \\
&& \bigotimes_{j=1}^{m} \big(|0\>_j+ \sigma_z^{(j-1)} |1\>_j \big) \, .
\end{eqnarray}
Here $\sigma_z^{(i)}$ is the Pauli matrix 
\begin{eqnarray}
\sigma_z^{(i)} \equiv |1\> \<1| - |0\> \<0|\, ,
\end{eqnarray}
with the exception of $\sigma_z^{(m)}$ and $\sigma_z^{(0)}$. These are given by 
\begin{eqnarray}
\sigma_z^{(m)}= \sigma_z^{(0)} \equiv |0\> \<0| + |1\> \<1| \, . 
\end{eqnarray}
By detecting odd parity for qubit $m+1$ and qubit $m$ we project these two qubits with $\proj_{01}+ \proj_{10}$. The resulting state is  
\begin{eqnarray}
|\psi \> &=& \frac{1}{2^{(n-1)/2}} \bigotimes_{i=m+2}^{n} \big(|0\>_i+ \sigma_z^{(i-1)} |1\>_i \big) \nonumber\\
&& \otimes \big(\sigma_z^{(m-1)} |0\>_{m+1}|1\>_m+ |1\>_{m+1}|0\>_m \big) \nonumber\\ 
&& \bigotimes_{i=1}^{m-1} \big(|0\>_i+ \sigma_z^{(i-1)} |1\>_i \big) \, . 
\end{eqnarray} 
The two originally independent chains are now linked via a double-encoded qubit, as illustrated in Fig.~\ref{fig:1d_fusion}. In order to remove the superfluous qubit we perform the Hadamard gate 
\begin{eqnarray} \label{H}
H =  {\textstyle \frac{1}{\sqrt{2}}} \big[ |0\> \<0| + |0\> \<1| + |1\> \<0| - |1\> \<1| \big]
\end{eqnarray}
on atom $m$ and measure its state. If qubit $m$ is found in $|0\>$, we need to apply a $\sigma_z$ operation to qubit $m-1$ to conclude the fusion of the two states. If we find qubit $m$ in $|1\>$, we instead perform the $\sigma_z$ operation on qubit $m-1$ and qubit $m+1$. In both cases we end up in the $n-1$ qubit cluster state 
\begin{eqnarray}
|\psi \> &=& \frac{1}{2^{(n-1)/2}} \bigotimes_{i=1}^{n-1} \big(|0\>_i + \sigma_z^{(i-1)} |1\>_i \big) \, .
\end{eqnarray}
Here we changed the indices of the atoms in order to close the gap caused by the loss of atom $m$. If instead the parity-check fails, and the projection $\proj_{00}$ or $\proj_{11}$ is performed, then the qubits $m+1$ and $m$ are both projected either on state $|0\>$ or $|1\>$. They are then decoupled from their respective cluster chains, which are now of length $n-m-1$ and $m-1$. In order to increase the efficiency of the growth of multi-qubit cluster states, one can abandon the measurement of atom $m$, as pointed out by Nielsen in Ref.~\cite{Nielsenxxx}. Its presence can be used later to increase the success rate for a later fusion of a cluster to this part of the chain.

\subsection{Fusion of two-dimensional clusters}

\begin{figure}[t]
\begin{minipage}{\columnwidth}
\begin{center}
\resizebox{\columnwidth}{!}{\rotatebox{0}{\includegraphics{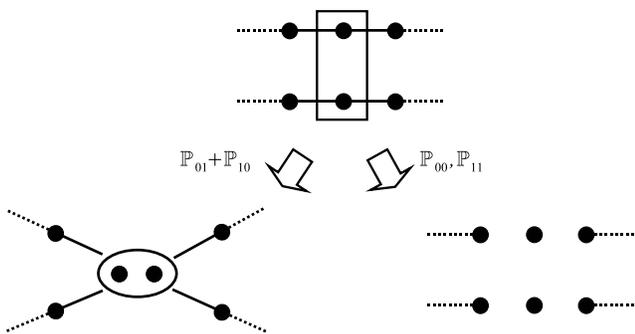}}}
\end{center}
\vspace*{-0.5cm} \caption{Fusion of two linear clusters into one two-dimensional cluster state. In case of a failure of the parity check, both chains split into shorter ones.}
\label{fig:2d_fusion}
\end{minipage}
\end{figure}

Similarly, large two-dimensional cluster states are obtained via the fusion of two smaller structures into one. As a simple example, we now discuss the fusion of two one-dimensional cluster chains of length $m$ and $n$ into a simple two dimensional structure, as illustrated in Fig.\ \ref{fig:2d_fusion}. The initial state of the two chains is given by 
\begin{eqnarray} 
|\psi \> &=& \frac{1}{2^{N/2}} \bigotimes_{i=1}^{n} \big(|0\>_i+ \sigma_z^{(i-1)} |1\>_i \big) \bigotimes_{j=1}^{m} \big(|0\>_j+ \sigma_z^{(j-1)} |1\>_j \big) \nonumber\\
&=& \dots \big(|0\>_k+ \sigma_z^{(k-1)} |1\>_k \big)  \dots  \big(|0\>_l+ \sigma_z^{(l-1)} |1\>_l \big)  \dots \nonumber \\
\end{eqnarray}
where $N=n+m$ is the total number of qubits.
For example, if the parity projection $\proj_{01} + \proj_{10}$ is successfully applied to qubit $k$ and qubit $l$, then these two qubits become a double encoded qubit which links the two chains. As a final step, one of the two atoms in the link, i.e.~atom $k$ or atom $l$, may be removed. In analogy to Section \ref{fusion}, this can be done using again a Hadamard operation and a qubit read out measurement followed by $\sigma_z$ operations. 

In the case of a $\proj_{00}$ or a $\proj_{11}$ projection, qubit $k$ and qubit $l$ decouple from the rest of the cluster states, since their state is now known. The original chains split and the fusion of the two chains failed. The situation is now  worse than before. Instead of one large cluster we obtained four smaller ones and two single qubits. Nevertheless, it is possible to grow cluster states of any size with the help of probabilistic measurements. This applies even when the probability for the successful fusion of two clusters is below $\frac{1}{2}$ \cite{Kok}. More details about the scaling behaviour of similar probabilistic cluster state growth schemes can be found for example in Refs.~\cite{Kok,Lim2,Nielsenxxx,Gross,Earl}. 

\section{Conclusions} \label{sec:conc}
 
In this paper we describe a scheme for the sequential build up of atomic cluster states with the help of the probabilistic parity measurement (\ref{eqn:parity}). This measurement can be implemented via the detection of a macroscopic fluorescence signal. It requires placing two atoms simultaneously into an optical cavity, where both experience comparable cavity-coupling constants and constant laser driving with comparable Rabi frequencies (c.f.~Fig.~\ref{fig:setup}). Fluorescence at a maximum level indicates that the atoms are in $|11 \>$, while fluorescence at a relatively low level indicates that the atoms are in $|01 \>$ or $|10 \>$, without revealing which atom is in which state. In the case of no cavity photon emissions, the atoms project into $| 00\>$. In Section \ref{sec:distinct}, we showed that the origin of these three distinct fluorescence levels is the existence of approximately decoupled subspaces in the effective evolution of the atomic ground states.  

One way to perform the parity measurement (\ref{eqn:parity}) is to turn on the laser field for a fixed time $T$ and to count the number of cavity photon detections in $(0,T)$. However, higher fidelities are achieved, when minimising the time $T$ for which the laser field is turned on. This minimises the effect of spontaneous emission from the atoms, which might disrupt the coherence between $|01 \>$ and $|10 \>$ or transfer population from $|11 \>$ into a state with one atom in $|0 \>$. We therefore propose an optimised protocol in Section \ref{sec:bitflip}, which makes use of the double heralding technique of Barrett and Kok \cite{Kok}. It requires to turn off the laser field upon the detection of the first photon and to swap of the states $|0 \>$ and $|1 \>$ in both atoms. Afterwards, another  laser pulse is applied for a maximum time $T_{\rm max}$ or until the detection of a second photon. In this way it is possible to measure how many atoms are in $|0 \>$ in a much shorter time than in the first mentioned protocol.

A detailed performance analysis of the optimised protocol can be found in Section \ref{sec:parity}. The main motivation for the proposed state preparation scheme is to allow for relatively large spontaneous decay rates and finite photon detector efficiencies $\eta$. Indeed, it is possible to achieve fidelities well above 0.99, when $\eta C \ge 20$, while $\eta C \ge 1$ is sufficient for fidelities above 0.88 (c.f.~Fig.~\ref{fig:parity_fidnumerics}). The success rate for an odd parity check is close to $\frac{1}{2}$ for most detector efficiencies $\eta$ and values of the single atom cooperativity parameter $C$ (c.f.~Fig.~\ref{fig:parity_prob_opt}). 
This means, the performance of the proposed state preparation scheme is essentially independent of the concrete size of the experimental parameters. Consequently the scheme is very robust against parameter fluctuations. To illustrate this, we show that  the fidelity reduces by only 0.1 even when the effective atom-cavity coupling strengths both differ by approximately $30 \, \%$ from their mean value. Fidelities in excess of 0.99 of the values calculated for equal coupling constants require that the atom-cavity couplings differ by less than $10 \, \%$ (c.f.~Fig.~\ref{fig:parity_fidrobust}).  

In Section \ref{sec:cluster}, we show how the parity measurement (\ref{eqn:parity}) can be used to grow two-dimensional cluster states. It has already been shown in the literature (c.f.~e.g.~Refs.~\cite{Kok,Lim2,Earl,Nielsenxxx,Gross}) that the build up of large cluster states is possible even when the probability for the successful fusion of two clusters is below $\frac{1}{2}$. Here we propose a scheme, in which the success rate for an odd parity check is close to $\frac{1}{2}$ even in the presence of finite efficiency photon detectors. Our cluster state growth scheme with macroscopic heralding is therefore expected to be much more practical than recent schemes based on the detection of single photons \cite{Cabrillo,Plenio,Kok,Lim,Lim2}. \\

\noindent {\em Acknowledgment.} 
We thank S. D. Barrett and P. L. Knight for stimulating discussions. A. B. acknowledges support from the Royal Society and the GCHQ. This work was supported in part by the EU Integrated Project SCALA, the EU Research and Training Network EMALI and the UK EPSRC through the QIP IRC.

\begin{appendix}
\section{Calculation of $P_n^{(L)}(t)$ and $P_n^{(H)}(t)$} \label{app:probs}

We now calculate the probability for $n$ atomic emissions from the $L$ to the $L$ subspace, given that the system is in this subspace at $t=0$ and remains there throughout. It is given by
\begin{eqnarray} \label{eqn:parity_PnL_app1}
P_n^{(L)}(t) &=& \int_{0}^{t} {\rm d}t_n \, \gamma_{LL} \, {\rm e}^{-(\gamma_{LL}+\gamma_{LD})(t-t_n)} \nonumber \\
&& \times \int_{0}^{t_n} {\rm d}t_{n-1} \, \gamma_{LL} \, {\rm e}^{-(\gamma_{LL}+\gamma_{LD})(t-t_{n-1})} \, ... \nonumber \\
&& \times \int_{0}^{t_2} {\rm d} t_1\, \gamma_{LL} \, {\rm e}^{-(\gamma_{LL}+\gamma_{LD})t_1} \, ,  
\end{eqnarray}
if the $t_i$ denote the corresponding jump times. The evaluation of the above integrals is straightforward and yields
\begin{eqnarray} \label{eqn:parity_PnL_app2}
P_n^{(L)}(t) &=& \frac{ \gamma_{LL}^n t^n}{n!} \, {\rm e}^{-(\gamma_{LL}+\gamma_{LD})t} \, .
\end{eqnarray}
Similarly, the probability for $n$ atomic emissions from the $H$ to the $H$ subspace, given that the system is in this subspace at $t=0$ and remains there throughout, is given by
\begin{eqnarray} \label{eqn:parity_PnH_app}
	P_n^{(H)}(t) &=&   
	\int_{0}^{t} {\rm d}t_n  ... \int_{0}^{t_2} {\rm d}t_1 \, \gamma_{HH}^n \, {\rm e}^{-(\gamma_{HH}+\gamma_{HL})t} \nonumber\\
	&=& \frac{ \gamma_{HH}^n t^n}{n!} \, {\rm e}^{-(\gamma_{HH}+\gamma_{HL})t}  \, .
\end{eqnarray}
\end{appendix}

\end{document}